\documentclass[twocolumn,showpacs,reprint,amsmath,amssymb,aps,prb]{revtex4-1}

\usepackage{graphicx}
\usepackage{dcolumn}
\usepackage{bm}
\usepackage{subfigure}
\usepackage{natbib}
\usepackage{multirow}
\usepackage{soul}
\usepackage[usenames,dvipsnames]{xcolor}
\usepackage{soul}
\usepackage{floatrow}

\begin{document}
\title{Thermoelectric properties of semimetals}
\author{Maxime Markov$^1$, Emad Rezaei$^1$, Safoura Nayeb Sadeghi$^2$, Keivan Esfarjani$^{2,3,4}$ and Mona Zebarjadi$^{1,4}$}

\email{m.zebarjadi@virginia.edu}

\affiliation{$^1$Department of Electrical and Computer Engineering, University of Virginia, Charlottesville, Virginia 22904, USA}
\affiliation{$^2$Department of Mechanical and Aerospace Engineering, University of Virginia, Charlottesville, Virginia 22904, USA}
\affiliation{$^3$Department of Physics, University of Virginia, Charlottesville, Virginia 22904, USA}
\affiliation{$^4$Department of Materials Science and Engineering, University of Virginia, Charlottesville, Virginia 22904, USA}

\begin{abstract}
Heavily doped semiconductors are by far the most studied class of materials for thermoelectric applications in the past several decades. They have Seebeck coefficient values which are 2-3 
orders of magnitude higher than metals, making them attractive for thermoelectric applications. Semimetals generally demonstrate smaller Seebeck coefficient values due to lack of 
an energy bandgap. However, when there is a large asymmetry in their electron and hole effective masses, semimetals could have large Seebeck coefficient values. In this work, we 
study the band structure of a class of 18 semimetals using first principles calculations and calculate their Seebeck coefficient using the linearized Boltzmann equation within the constant relaxation time approximation. We conclude, despite the 
absence of the band gap, that some semimetals are good thermoelectrics with Seebeck coefficient values reaching up to 200 $\mu$V/K. We analyze the metrics often used to describe 
thermoelectric properties of materials, and show that the effective mass ratio is a key parameter resulting in high Seebeck coefficient values in semimetals.
\end{abstract}


\maketitle

\section{Introduction.}

Material thermoelectric figure of merit, $zT$ is defined as $zT = \frac{\sigma S^2T}{\kappa}$, wherein $\sigma$ is the electrical conductivity, $S$ is the Seebeck coefficient, $\kappa$ is the 
thermal conductivity and T is the temperature. A thermoelectric module is made out of n and p legs electrically in series and thermally in parallel. The efficiency of a thermoelectric module in power generation mode~\cite{Goldsmid:2001} and in refrigeration cycle~\cite{Ioffe:1957}, and the thermal switching ratio in switching mode~\cite{Adams:2019} are increasing functions of the n and p materials' figure of merit. Hence, finding thermoelectric materials with large figure of merit is of high interest. 

Metals were the first class of materials studied for thermoelectric applications. While they have large electrical conductivity, they usually have small Seebeck coefficient values and large thermal conductivity values, making them non ideal candidates for traditional thermoelectric applications. Semiconductors usually own Seebeck coefficient values that are orders of magnitude larger than metals. The large Seebeck coefficient is the result of the presence of the bandgap which breaks the symmetry between electrons and holes. There are two major competing factors in optimization of the figure of merit in semiconductors. First, when the Fermi level is inside the bandgap, the Seebeck coefficient is large. As the Fermi level moves into the valence or conduction bands, the difference between the density of states (DOS) of hot electrons (above the Fermi level) and cold electrons (below the Fermi level) becomes small, and so does the Seebeck coefficient. 
On the contrary, the electrical conductivity increases since there are more available electronic states. As a result, one needs to adjust the position of the Fermi 
level to optimize the thermoelectric power factor, $P = \sigma S^2 $. Second, as the Fermi level moves inside the band, similar to the electrical conductivity, the electronic part of the thermal conductivity also increases. It is therefore difficult to design a material with very large figure of merit although no theoretical upper limit has been found for $zT$. 

Semimetals are a class of materials with properties in between semiconductors and metals. They usually have very small overlap of bands and therefore while they do not have an energy gap, their intrinsic carrier density can vary in a large range, between $10^{15} - 10^{20}$ cm$^{-3}$, depending on the band overlap and the size of the carrier pockets. For example, the intrinsic concentrations at liquid helium temperature 4.2 K are about $5.0\cdot10^{15}$ cm$^{-3}$ for HgTe~\cite{Guldner:1973}, $3.6\cdot10^{16}$ cm$^{-3}$ for HgSe~\cite{Lehoczky:1974}, $2.7\cdot 10^{17}$ cm$^{-3}$ for Bi~\cite{Issi:1979}, $3.7\cdot10^{19}$ cm$^{-3}$ for Sb~\cite{Issi:1979}, and $2.0\cdot10^{20}$ cm$^{-3}$ for As~\cite{Issi:1979}. These values are much smaller than in metals, which are typically around $10^{23}$ cm$^{-3}$, and are comparable with and in some cases smaller than in heavily-doped semiconductors used for thermoelectric applications, $10^{18} - 10^{20}$ cm$^{-3}$. 
However, semimetals generally have much larger carrier mobility values compared to metals and heavily doped semiconductors. For example, electron mobilities at 4.2 K are $6.0\cdot 10^5$
$cm^2 V^{-1}s^{-1} $ in HgTe~\cite{Guldner:1973}, $1.2\cdot10^{5}$ $cm^2 V^{-1}s^{-1}$ in HgSe~\cite{Berger:1997,Nikolskaya:1955} and $11\cdot 10^{7}$ $cm^2 V^{-1}s^{-1}$ in Bi~\cite{Issi:1979}. As a result, the electrical conductivity of semimetals 
is comparable to those of heavily-doped semiconductors. Note that the carrier mobility is much lower in heavily-doped semiconductors due to ionized impurity doping and in metals due 
to electron-electron and electron-phonon interactions. The thermal conductivity values in semimetals could be also small, especially if they consist of heavy elements. For example, the thermal 
conductivity at room temperature is about 1.7 $W m^{-1}K^{-1}$ in HgSe~\cite{Slack:1972}, 1.9-2.9 $W m^{-1}K^{-1}$ in HgTe~\cite{Slack:1972,Markov:2018}, 6.0 $W m^{-1}K^{-1}$ in the trigonal direction in pure 
bismuth~\cite{Gallo:1963,Lee:2014,Markov:2016} and could be as low as 1.6 $W m^{-1}K^{-1}$ in Bi-Sb alloys~\cite{Kagan:1991,Lee:2014}. 

Semimetallic and zero gap materials show many interesting properties. They have attracted interests as topologically non trivial materials~\cite{Singh:2018}. Many of them have strong spin-orbit coupling and comprise of heavy elements. As a result, they possess a low thermal conductivity. Inversion of bands happens in many of the zero-gap alloys such as Bi$_x$Sb$_{1-x}$ and Hg$_x$Cd$_{1-x}$Te, leading to interesting transport properties. While many of these materials have been studied in other fields, there has not been a systematic study of their thermoelectric properties due to their lack of band gap.

If one is to avoid doping and only choose to work with intrinsic materials, semimetals would be the best potential candidate for having a large thermoelectric power factor~\cite{Zebarjadi:2018}. 
In our recent publication, we studied thermoelectric properties of HgTe~\cite{Markov:2018} as a well-known semimetal. One of the interesting features observed was that the Seebeck coefficient in 
intrinsic HgTe was not sensitive to the number of defects and impurities inside the sample. This means one can change the carrier concentration by orders of magnitude while keeping 
the Seebeck coefficient constant. This is because these large changes in carrier concentration did not result in a considerable shift in the chemical potential, so that Seebeck was not changed. If this is a general trend in semimetals, then the interplay between electrical conductivity and Seebeck coefficient is much weaker in semimetals compared to semiconductors and therefore it is easier to increase the figure of merit in semimetallic samples especially at low temperatures where the dominant source of scattering is impurity scattering. 

\begin{figure}[t]
    \centering
    \includegraphics[width=1.0\linewidth]{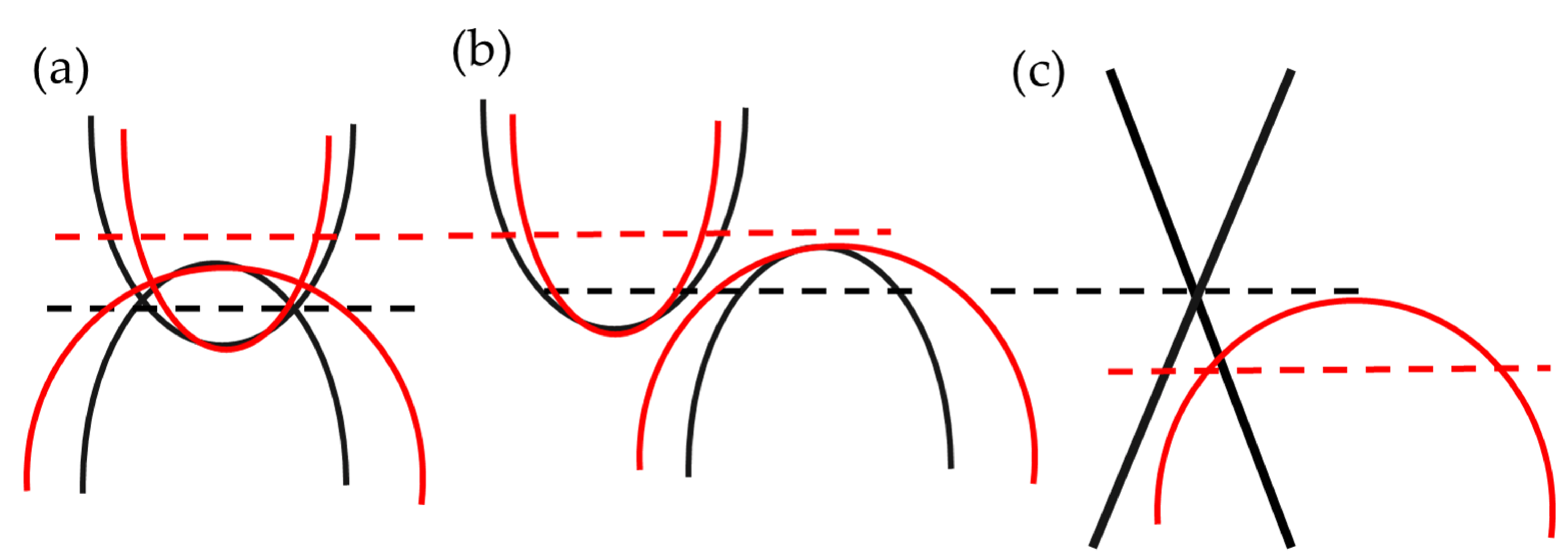}
    \caption{Schematic illustration of different types of semimetals: a) direct semi-metal with parabolic bands b) indirect semimetal with parabolic bands c) Dirac or Weyl semimetal with linear dispersion. The Fermi level is denoted by the dashed line. In each case, the band structure could be symmetric as shown by black curves or asymmetric as shown by red curves. Semi-metals with asymmetric bands are the focus of this work}
    \label{fig:sketch-semimetals}
\end{figure}

In this work, the thermoelectric response of several semimetallic elements, i.e. their Seebeck coefficient values, are studied using first principles calculations with proper corrections for the 
energy levels. We restrict ourselves to room temperature where the diffusive part of the Seebeck coefficient is known to be dominant.
The rationale to focus only on the Seebeck coefficient is the following: As was discussed, the carrier mobility is expected to be large in semimetals.
If semimetals consist of heavy elements, then their thermal conductivity is also expected to be low. The biggest concern with semimetals is therefore the Seebeck coefficient and thus the process of searching for good semimetals for thermoelectric applications should start with the scan for the Seebeck coefficient. From a computational point of view, 
among the three transport properties determining the figure of merit, the Seebeck coefficient is the least sensitive one to the scattering rates. Therefore, the only property that 
could be reliably calculated under constant relaxation time approximation and still be of value, is the Seebeck coefficient. We should acknowledge that even Seebeck coefficient values can be 
modified when energy dependent relaxation times are introduced~\cite{Samsonidze:2018, Ahmad:2010, Markov:2018, Pshenay-Severin:2018}. Including energy dependent relaxation times would be a very difficult task when scanning many materials. Here, as the first step towards finding promising semimetallic candidates, we limit ourselves to the constant relaxation time approximation.

\begin{table}[t]
 \centering
 \begin{tabular}{ccccccc}
    \hline  
    \hline
         & \hspace{0.1cm}  & Irr. k-points & \hspace{1cm}& cut-off energy & \hspace{0.3cm} &  BTE solver\\
    & \hspace{0.2cm}  & \scriptsize{PBE, HSE/ mBJ} & \hspace{0.4cm}& \scriptsize{PBE, HSE/ mBJ} \\
    \hline 
    \hline 
        &&&&&&\\
HgTe          &              & 1661, 752      &     &          350, 350    &               &   BoltzTraP1 \\
HgSe          &              & 1661, 752      &     &          420, 420    &               &   BoltzTraP1 \\
$\alpha-$HgS  &              & 1661, 752      &     &          420, 420    &               &   BoltzTraP1 \\
TlP           &              &  413, 321      &     &          420, 420    &               &   BoltzTraP1 \\
TlAs          &              &  413, 321      &     &          420, 420    &               &   BoltzTraP1 \\
Li$_2$AgSb    &              &  413, 104      &     &          420, 420    &               &   BoltzTraP1 \\
Na$_2$AgSb    &              &  413, 104      &     &          420, 420    &               &   BoltzTraP1 \\
Rb$_2$AgSb    &              &  413, 104      &     &          420, 420    &               &   BoltzTraP1 \\
$\alpha-$Sn   &              &  294, 294      &     &          600, 560    &               &   BoltzTraP2 \\
Sb            &              &  868, 868      &     &          600, 500    &               &   BoltzTraP2 \\
Bi            &              &  868, 868     &     &          600, 560    &               &   BoltzTraP2 \\
TaAs          &              &  512, 512     &     &          600, 500    &               &   BoltzTraP2 \\
TaP           &              &  512, 512     &     &          600, 500    &               &   BoltzTraP2 \\
NbP           &              &  512, 512     &     &          600, 500    &               &   BoltzTraP2 \\
Mg$_2$Pb      &              &  294, 294     &     &          600, 560    &               &   BoltzTraP2 \\
PtSb$_2$      &              &  216, 216     &     &          600, 600    &               &   BoltzWann \\
TiS$_2$       &              &  667, 108     &     &          500, 500    &               &   BoltzWann \\
TiSe$_2$      &              &  667, 108     &     &          500, 500    &               &   BoltzWann \\
    \hline  
    \hline 
 \end{tabular}
 \caption{\textbf{\label{tab:comp-details}}. Computational details of DFT calculations including number of irreducible k points in the Brillouin zone, cut-off energy (eV) for each material and used XC functional as well as BTE solver code}
 \end{table}

\begin{figure*}[t]
\centering
  \subfigure[][]{\includegraphics[width=0.3\linewidth]{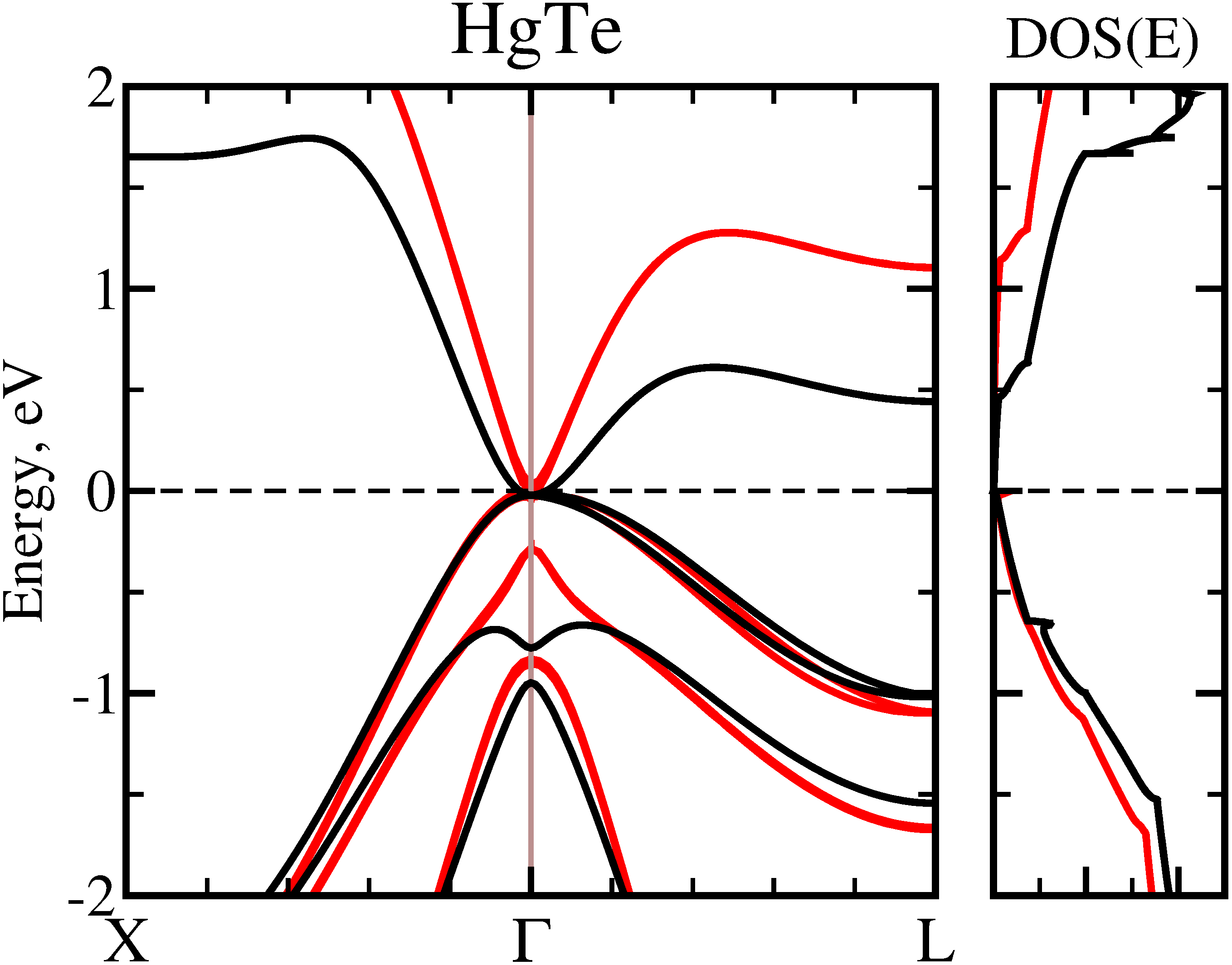}}
~
  \subfigure[][]{\includegraphics[width=0.3\linewidth]{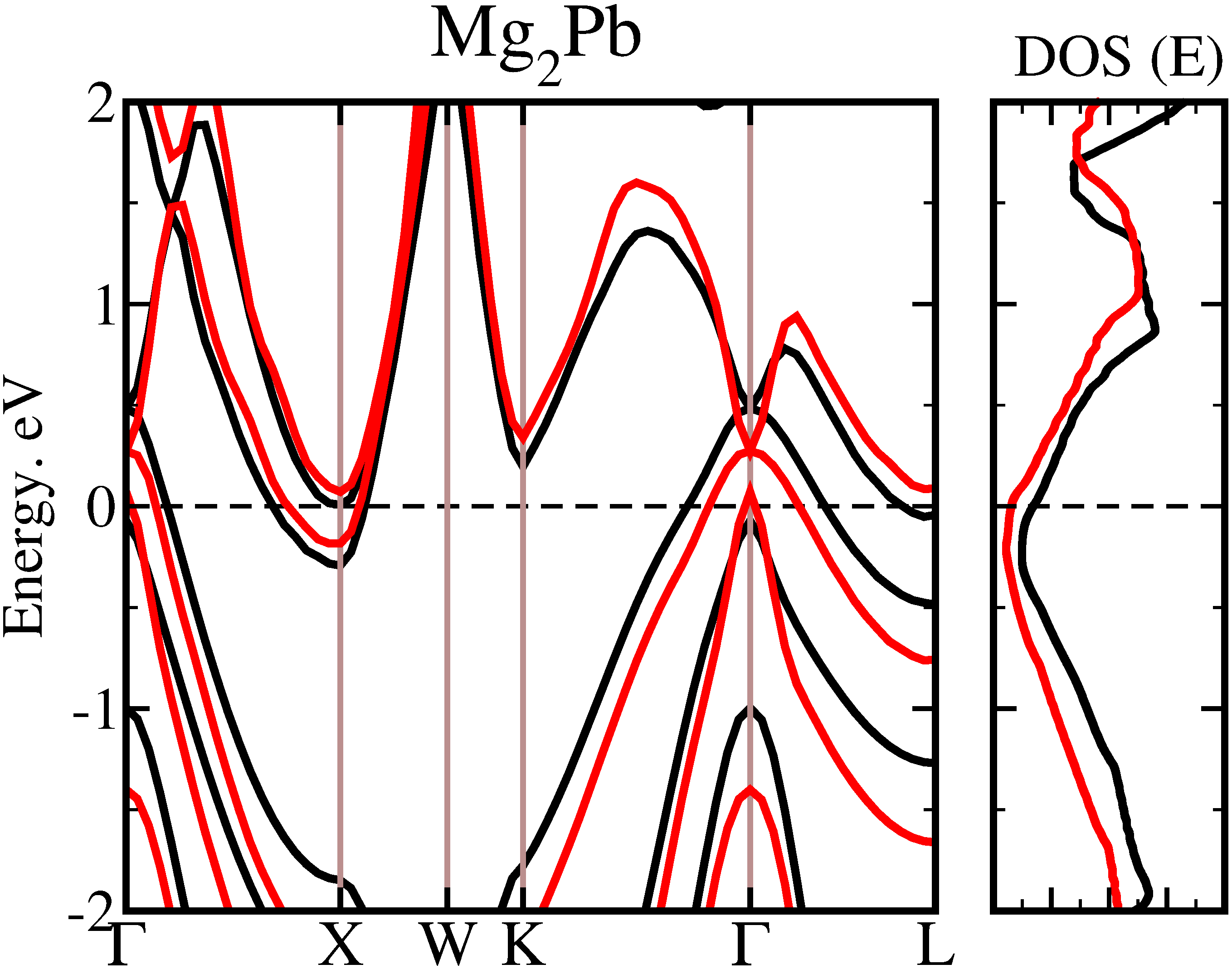}}
~
  \subfigure[][]{\includegraphics[width=0.3\linewidth]{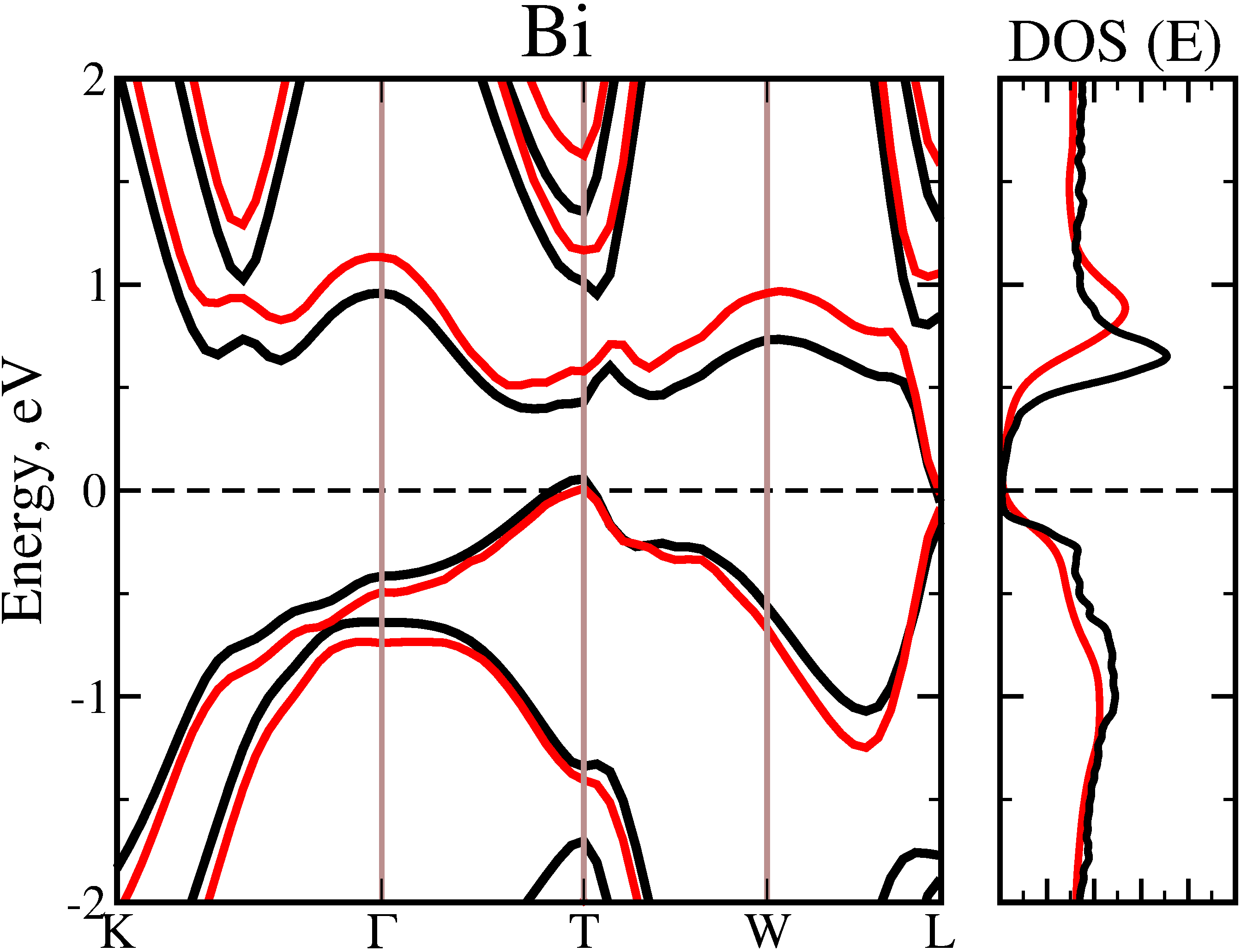}}
   \caption{The band structures (left panels) and density of states (right panels) of HgTe, Mg$_2$Pb and Bi representing the three types of semi-metals described in Figure~\ref{fig:sketch-semimetals}. Black curves - PBE, red curves - HSE06.}
   \label{fig:3mat}
\end{figure*}

\section{Computational Methods}

We preselect 18 materials which were mentioned in the literature as semimetals. Information about their crystal structure, space group number and number of atoms 
per unit cell are summarized in Table~\ref{tab:material}. Vienna Ab-initio Simulation Package (VASP)~\cite{Kresse:1996,Kresse:1996a} is used to perform first-principles calculations. Pseudopotentials based on the projector augmented wave method~\cite{Blochl:1994} from VASP library with the generalized gradient approximation by Perdew, Burke and Ernzehof 
(GGA-PBE)~\cite{Perdew:1996} as well as hybrid Heyd-Scuseria-Ernzehof (HSE06)~\cite{Heyd:2003} exchange-correlation (XC) functionals are employed to calculate band structure and density of states. The details about the cut-off energy and number of k points may be found in Table~\ref{tab:comp-details}. We used relaxed PBE lattice parameters for all materials but for HgTe, HgSe and HgS. For the latter three materials the experimental lattice parameters~\cite{Fleszar:2005,Sakuma:2011} were considered. The summary of the lattice parameters can be found in the Supplementary Material. Spin-orbit coupling is included in all calculations (except for TiS$_2$ and TiSe$_2$) and transport calculations are performed within the Constant Relaxation Time Approximation (CRTA) as implemented in BoltzTraP~\cite{Madsen:2006, Madsen:2018} and BoltzWann~\cite{Pizzi:2014} codes to obtain the diffusive part of the Seebeck coefficient (see Table~\ref{tab:comp-details}). The interpolating k point grid was taken to be at least 30 times denser than the initial DFT grid.

\section{Results}

We divide all semimetals into three separate groups. These are shown schematically in Fig.~\ref{fig:sketch-semimetals}. The first group possesses a distinct feature in the band structure: the lowest conduction band has a deep minimum at the center of the Brillouin zone (BZ) where it overlaps with the highest valence bands. When the two bands are symmetric (shown by black curves), the intrinsic chemical potential is expected to be at the midpoint between the to band extrema, and the intrinsic Seebeck coefficient is expected to be very small. However, it is possible to have a band structure similar to the red curve in Fig~\ref{fig:sketch-semimetals}a, where the low degeneracy of the conduction band in vicinity of the $\Gamma$ point results in a small density-of-states (DOS), the magnitude of which is essentially defined by the electron's effective mass (\textit{i.e.} the curvature of the band). On the other hand, valence bands have heavier effective masses and higher degeneracy with contributions from elsewhere in the BZ. As a result, the DOS is asymmetric about the chemical potential. This is known to be beneficial for the material's electronic properties in general and, in particular, leads to a high Seebeck coefficient. A typical example of such material is HgTe which has been studied in our recent publication both theoretically and experimentally~\cite{Markov:2018}. Other (predominantly cubic) materials are HgSe and HgS, TlAs and TlP~\cite{Lin:2013}, $\alpha$-Sn as well as inverse Heusler materials (Li$_2$AgSb, Na$_2$AgSb, Rb$_2$AgSb)~\cite{Lin:2013,Li:2014}. The band structures of these materials along with their DOS are shown in the Supplementary Material. Here, as the representative of this class of materials, we show the band structure and the DOS of HgTe as shown in Fig.~\ref{fig:3mat}a. Black curves are used to show PBE results for the band structure and the DOS of all materials reported in this work. Red curves show the HSE results.

Among the materials studied within the first class, the hybrid functional calculations (red curves) reveal that HgS and Li$_2$AgSb are in fact semiconductors with band gaps of 0.33~eV and 0.67~eV respectively. In almost all cases, we note that the effective masses of the conduction band significantly decreases in HSE comparing with PBE calculations. A possible explanation for this effect has been given in Ref.~\cite{Kim:2009} where the small effective masses were attributed to the strong level repulsion between the s-like conduction band and p-like valence band at $\Gamma$. This repulsion is inversely proportional to the square of the difference between these two levels~\cite{Kim:2009} which reduces from $-0.93$ eV for PBE to -0.27 eV for HSE06 in case of HgTe~\cite{Markov:2018,Nicklas:2011}

\begin{figure*}[t]
    \centering
    \includegraphics[width=0.70\textwidth]{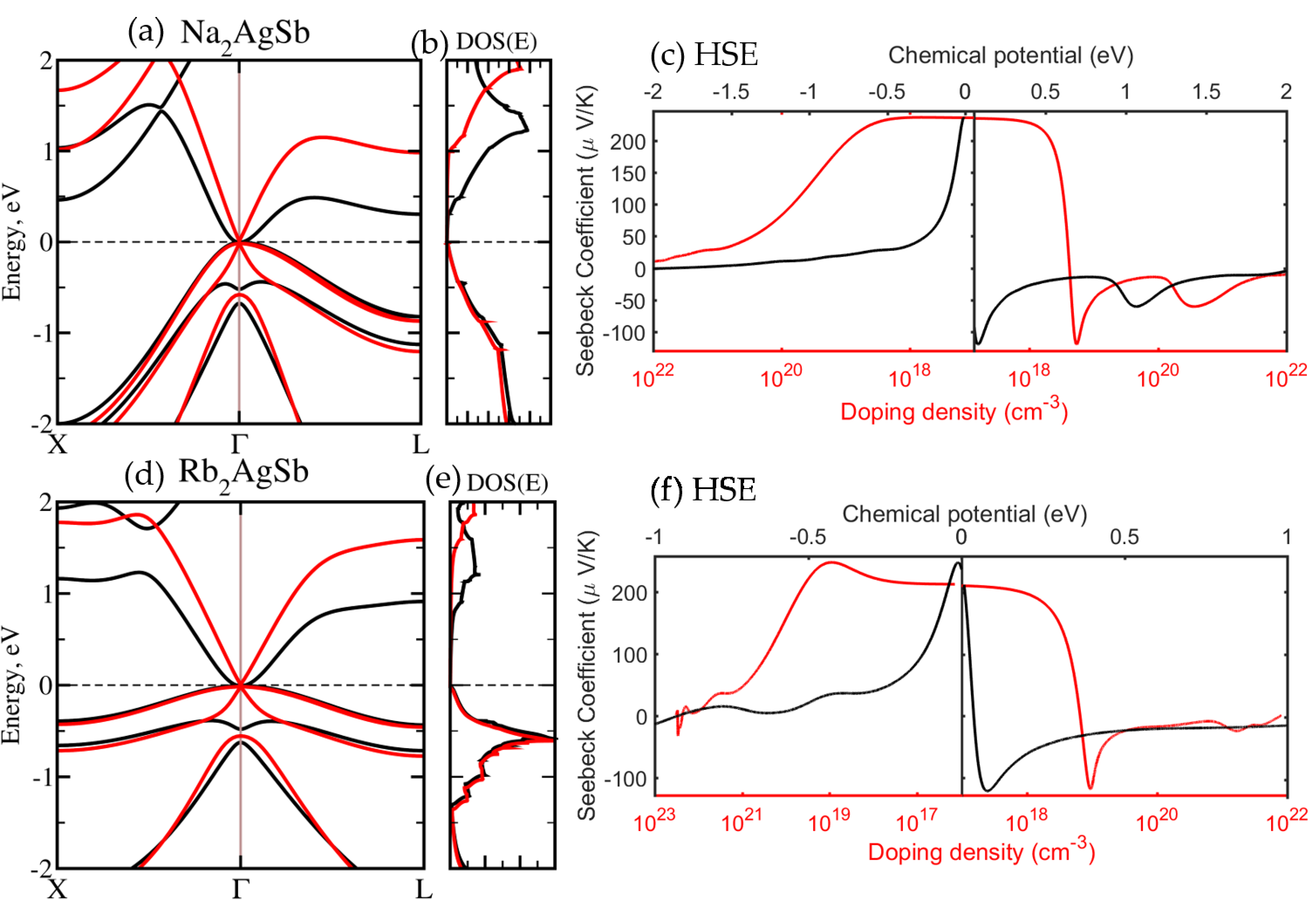}
    \caption{The band structure, the density of states and the Seebeck coefficient of Na$_2$AgSb are shown in panels a, b, and c respectively. Similarly those of Rb$_2$AgSb are shown in the lower panels of d,e, and f. The black curves in the band structure and DOS plots are PBE results and the red curves are HSE results. The Seebeck coefficients are only reported for HSE results. Red curves are Seebeck coefficient values vs. doping concentration and black curves are vs. chemical potential. Left side of each Seebeck plot refers to p-type doping and right side to n-type doing.}
    \label{fig:BestS}
\end{figure*}

The second group (Fig.~\ref{fig:BestS}b) includes other semimetallic materials without any distinct feature in their band structure but possessing a low density of states at the Fermi level. The top of the valence band and the bottom of the conduction band are at different k points as shown schematically in Fig.~\ref{fig:sketch-semimetals}b. Electron and hole pockets coexist. This class includes, for instance, Mg$_2$Pb, cubic pyrite structures (PtSb$_2$ and PtBi$_2$)~\cite{Saeed:2013,Gao:2017,Sondergaard:2012}, 
TiS$_2$, TiSe$_2$, TaP, NbP and $\alpha-$Zn$_3$Sb$_2$.~\cite{Lo:2017} We note TiS$_2$ gap opens up when HSE functional is used and therefore this material is a semiconductor with the bandgap of 0.4~eV. Despite its large Seebeck coefficient which is expected for a material with a bandgap, the intrinsic carrier concentration is low and therefore it does not fall in the class of materials we are interested in this work. On the other hand, TiSe$_2$ remains semimetallic under HSE, with overlapping conduction (L and M points) and valence bands ($\Gamma$ point). Its Seebeck coefficient is however found to be small due to the small asymmetry in the bands. In another work where properties of the monolayer TiSe$_2$ were studied\cite{Sadeghi:2019}, we found that the bandgap can be opened under tensile strain, leading to a metal-insulator transition and corresponding non-linear effects. 
As for Mg$_2$Pb, the overlap of bands is relatively large and the ratio of the DOS effective mass of the conduction band to that of the valence band is close to one. (see Fig.~\ref{fig:BestS}b) Therefore this material exhibits a small intrinsic Seebeck coefficient value of about -10$\mu V/K$. 

\begin{figure}[b]
    \centering
    \includegraphics[width=1.0\linewidth]{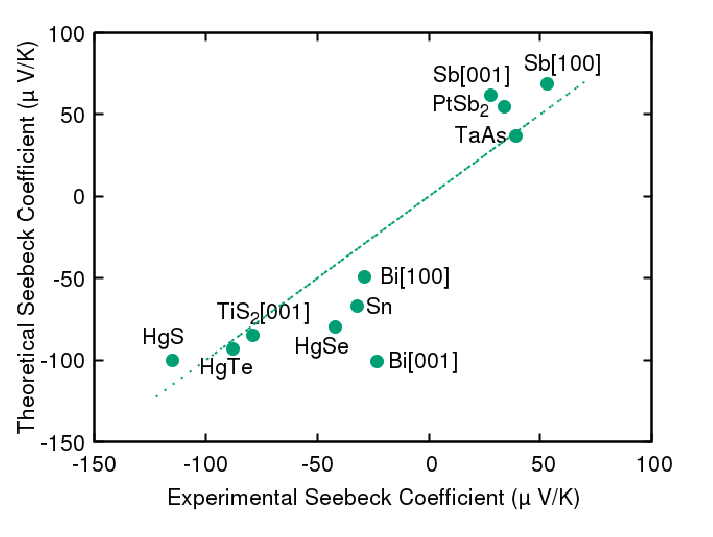}
    \caption{Computational Seebeck coefficient values calculated in this work using HSE band structures versus experimental Seebeck coefficient values from literature. When available, single crystals in [001] (trigonal) and [100] (binary) directions are compared. Most experimental samples are intrinsic including  Bi~\cite{Rowe:2006}, Sb~\cite{Saunders:1965} ,Sn~\cite{Goland:1956}, HgTe~\cite{Whitsett:1972}, PtSb$_2$~\cite{Waldrop:2015} , and TaAs~\cite{Saparov:2012}. Other samples are n-doped including:  $\alpha$-HgS~\cite{Shchennikov:2007} ($10^{18}$cm$^{-3}$), TiS$_2$~\cite{Wang:2018} ($8\cdot10^{18}$cm$^{-3}$), and HgSe~\cite{Ramesh:1982}( $10^{18}$cm$^{-3}$).}
    \label{fig:CompareS}
\end{figure}

\begin{figure*}[t]
    \centering
    \includegraphics[width=1.0\linewidth]{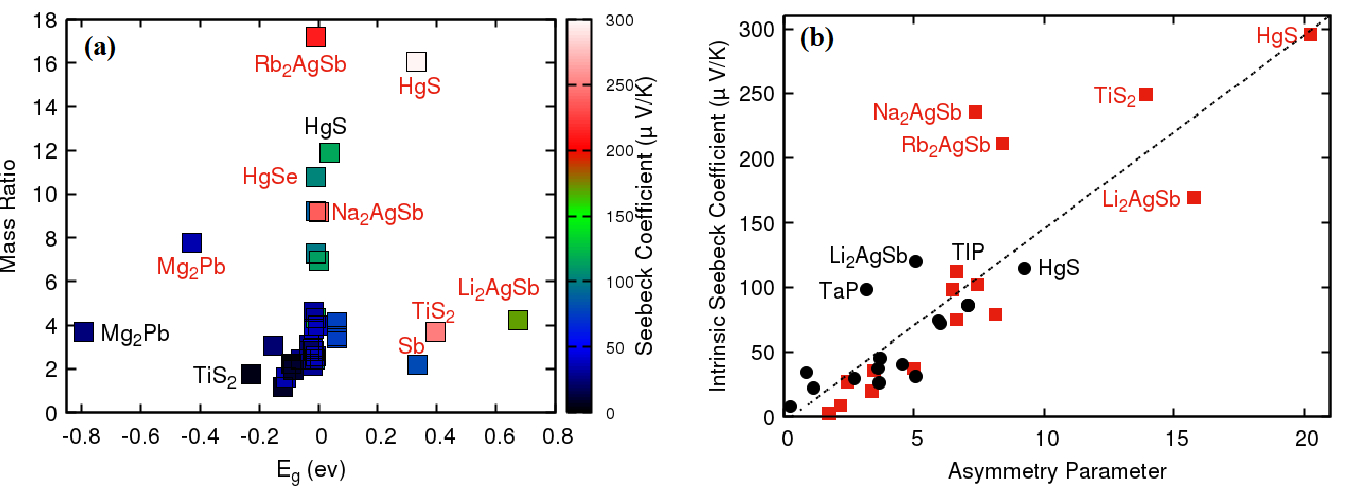}
    \caption{(a) Absolute value of intrinsic Seebeck coefficient (color bar) as a function of band gap $E_g$ (x-axis) and effective mass ratio $\gamma$ (y-axis). Negative bandgap refers to overlapping bands. (b) Intrinsic Seebeck coefficient as a function of asymmetry parameter $\Xi$ defined in eq.~\ref{eq3}. Black circles - PBE data, red squares - HSE06 data, black dotted line is a linear dependence assumed from Eq.~\ref{eq2}}
    \label{fig:ASYM}
\end{figure*}

The third class of materials includes relativistic (Dirac and Weyl) semimetals with linear bands close to the Fermi level. These are schematically shown in Fig.~\ref{fig:sketch-semimetals}c. The examples include Bi, Sb, Na$_{3}$Bi and TaAs-family and inverse Heusler materials Na$_2$AgSb and Rb$_2$AgSb. Thermoelectric properties of the latter family as well as some other topologically non-trivial semimetals have been recently investigated in Ref.~\cite{Singh:2018}. The band structure of Bi as the representative of this class of materials is shown in Fig.~\ref{fig:BestS}c. Most samples in this group demonstrate rather small Seebeck coefficient values.   
This is expected because there is an inherent symmetry in the band structure at the Dirac point. The symmetry can breakdown only if additional bands exist close to the Dirac point as shown schematically in Fig.~\ref{fig:sketch-semimetals}c. 

Two examples are Na$_2$AgSb and Rb$_2$AgSb which show, within HSE, a Dirac dispersion at $\Gamma$ point, in addition to a parabolic valence band. The band structure, the DOS and the corresponding Seebeck coefficient of these two materials are shown in Fig.3. The Seebeck is only reported for HSE calculations. For PBE results, where bands were parabolic instead of linear, we refer the reader to the supplementary materials. Both materials show large intrinsic Seebeck coefficient values and large intrinsic carrier concentrations.  The Seebeck coefficient of Na$_2$AgSb is larger than 200 $\mu V/K$ and interestingly it is insensitive to the changes in the carrier concentration up to $\pm$ 10$^{18}$ cm$^{-3}$. The large value can be associated with the extra parabolic valence band, and the flat Seebeck to the constancy of the slope of the DOS and group velocities in this region. 
Rb$_2$AgSb is similar. We see this trend more or less for all our calculated materials, indicating that the coupling between electrical conductivity and the Seebeck coefficient is weaker in the semimetallic samples compared to heavily doped semiconductors. We also emphasize that these large Seebeck values are obtained at relatively large carrier concentrations. Note the carrier concentration reported in the plots, are the Hall type carrier concentration, i.e., the difference between free electron, n, and free holes, p, densities. The actual carrier concentration that determines the electrical conductivity is larger and is the sum of n and p. 

Bismuth and antimony are well-known materials and have been the subject of studies for many years. The experimentally measured values for Bi and Sb can provide a good comparison to the theoretical calculations. In addition to Bi and Sb, in Fig.~\ref{fig:CompareS} our computational results are compared to reported experimental values of $\alpha$-Sn, $\alpha$-HgS, HgSe, HgTe, TiS$_{2}$, TaAs and PtSb$_{2}$.As shown in Fig.~\ref{fig:CompareS}, and considering there are no fitting parameters, the agreement between theory and experiment is satisfactory.   

Several of the samples that we have studied in this work have Seebeck coefficient values larger than 100$\mu V/K$. We expect the Seebeck coefficient to be large only when there is a band gap or when there is asymmetry between electron and hole effective masses. To demonstrate this, we extract an effective mass from the density of states estimated as the slope of density of states of the electrons (conduction band) and the holes (valence band) with respect to the square root of energy. The absolute value of intrinsic Seebeck coefficient of different materials with respect to the effective mass ratio (effective mass of the holes to that of the electrons) and bandgap energy is plotted in Fig.~\ref{fig:ASYM}a. We see an increasing trend in the Seebeck coefficient values with respect to the mass ratio for semimetals where the band gap is zero or close to zero. There are a few exceptions, namely Na$_2$AgSb and Rb$_2$AgSb. In these materials, due to the presence of the Dirac point, the parabolic assumption enabling the extraction of an effective mass is not accurate. 

There are two parameters to which Seebeck is sensitive: one is the bandgap and the position of the chemical potential, and the other is the mass ratio.  
For this reason, we would like to define a {\it single} parameter which we call "asymmetry parameter" to characterize Seebeck, and show their correlation.

We start by using the equation for bipolar Seebeck coefficient under constant relaxation time approximation~\cite{Mahan:1998}
\begin{equation}
S =- \frac{k_B}{2q} \left[\frac{\sigma _e-\sigma_h}{\sigma_e+\sigma_h} (\beta E_g+5)+\beta (\varepsilon_c+\varepsilon_v-2\mu)\right] 
\label{eq1}
\end{equation}
where k$_{B}$ is the Boltzmann constant, q is the elementary charge, $\beta = (k_{B} T)^{-1}$,  $\sigma_{e}$ and $\sigma_{h}$ are electron and hole conductivities. The bandgap $E_{g}$ is defined similar to semiconductors as a difference between the bottom of the conduction band $\varepsilon_{c}$ and the top of the valence band $\varepsilon_{v}$ with the the chemical potential $\mu$ somewhere in between. The bandgap $E_g$ is positive for semiconductors and  negative for semimetals where there is band overlap. Its values for different materials studied in this work are listed in Table~\ref{tab:material}.

\begin{table*}[t]
 \centering
 \begin{tabular}{ccccrrrrrrrr}
    \hline
    \hline
                   &         &         &             &  &                        &    &                 &  &   &&    \\    
        Material   & Crystal & Group\# & $N_{atoms}$ &  &  $E_{g}$ (eV)          &    & $E_{g}$ (eV)    &  & $m^{*}_{h}/m^{*}_{e}$ & & $m^{*}_{h}/m^{*}_{e}$\\
                   &         &         &             &  &  \textbf{PBE}          &    &  \textbf{HSE/mBJ} &  &  \textbf{PBE} &  & \textbf{HSE/mBJ}\\ 
                   &         &         &             &  &                        &    &                 &  &   &   & \\ 
    \hline
    \hline
    \textbf{HgTe}  &  cubic  &  216    &    2         &  & -0.019              &    &  -0.009  &  &2.14 && 7.25 \\
    \textbf{HgSe}  &  cubic  &  216    &    2         &  & -0.018              &    &  -0.009  &  & 3.85 && 10.78 \\
    \textbf{HgS}   &  cubic  &  216    &    2         &  & 0.038              &    &   0.305   &  & 11.90 && 16.06 \\    
    \textbf{TlP}   &  cubic  &  216    &    2         &  & -0.018              &    &  0.000     &  & 2.84 && 6.95 \\
    \textbf{TlAs}  &  cubic  &  216    &    2         &  & -0.019              &    &  -0.009     &  & 2.93 && 2.57 \\
\textbf{Li$_2$AgSb}&  cubic  &  216    &    2         &  & -0.009              &    & 0.676 &  & 4.30 && 2.62 \\
\textbf{Na$_2$AgSb}&  cubic  &  216    &    2         &  & -0.009              &    &  0.000     &  & 6.20 && 9.19 \\
\textbf{Rb$_2$AgSb}&  cubic  &  216    &    2         &  & -0.020              &    & -0.008      &  & 8.61 && 17.18 \\
\textbf{$\alpha$-Sn}& cubic  &  227    &    2         &  & -0.031        &    &   -0.030     &  & 3.11 && 2.83 \\
   \textbf{Bi}     & trigonal&  166    &    2         &  & -0.122           &    & -0.061      &  &1.15 && 2.44 \\
   \textbf{Sb}     & trigonal&  166    &    2         &  & 0.061     &    &0.335               &  &3.49 && 2.21 \\
\textbf{TaAs}                  &  tetragonal   &  109  & 4   &  &0.062    &    & -0.003   &  & 0.29 && 0.25\\  
\textbf{TaP}                   &  tetragonal   &  109  & 4   &  & -0.15   &    & -0.092  &  & 0.41  && 0.44\\  
\textbf{NbP}                   &  tetragonal   &  109  & 4   &  &   -0.152   &    &  0.061  &  & 0.33 && 0.24\\  
  \textbf{Mg$_2$Pb}            &  cubic        &  225  & 3   &  & -0.793         &    & -0.427 &  & 0.27 && 0.13\\  
\textbf{PtSb$_2$}              &  cubic        &  205  & 12  &  & -0.110          &   & -0.083    & & 1.61 && 1.94\\     
\textbf{TiS$_2$}              &  trigonal     &  164  &  3  &  &-0.226         &    & 0.396  &  & 0.57 && 0.27\\ 
\textbf{TiSe$_2$}              &  trigonal     &  164  &  3  &  &-0.623         &    & -0.346  &  & N/A && N/A\\
    \hline  
    \hline 

 \end{tabular}
 \caption{\label{tab:material} Summary of 18 materials studied in this work including their crystal structure, space group number, number of atoms per unit cell, band gap in PBE and HSE 
 calculations as well as ratio of hole effective mass to electron effective mass.}
\end{table*}

Assuming non-degenerate statistics, constant relaxation time approximation and intrinsic conditions ($n=p$), one can simplify Eq.~\ref{eq1} to
\begin{equation}
S =-\frac{k_B}{2q} \left[\frac{\gamma-1}{\gamma+1} (\beta E_g+5)-\frac{3}{2} \ln(\gamma)\right]
\label{eq2}
\end{equation}
where $\gamma$ is defined as effective mass ratio of holes to electrons (listed in Table~\ref{tab:material}). Note that the condition $n=p$ automatically places the chemical potential at the right place, and we do not need to specify it.
Since we are interested in the absolute value of the Seebeck coefficient, we define the following parameter as the indicator of asymmetry: 
\begin{equation}
\Xi =\left|\frac{\gamma -1}{\gamma+1}\right| (\beta E_g+5)+\left|\frac{3}{2} ln(\gamma)\right|
\label{eq3}
\end{equation}

In Figure~\ref{fig:ASYM}b, we show the dependence of the calculated intrinsic Seebeck coefficient to the asymmetry parameter $\Xi$. According to Eq.~\ref{eq2}, a linear dependence to asymmetry parameter is expected. The obtained results show a noisy linear trend. Therefore, we conclude that this parameter can be used to estimate the Seebeck coefficient from the band structure information. The fact that there is a large level of noise is attributed to several factors. First, many of the studied materials do not obey the parabolic band dispersion and therefore it is not possible to define a proper effective mass for them. Second, the model uses non-degenerate statistics that is not accurate when there is an overlap between the bands. Despite these, there is clear increasing trend of the Seebeck coefficient with respect to the defined asymmetric parameter.

\section*{Acknowledgements}
This work is support by National Science Foundation, grant number 1653268. Calculations were performed using Rivanna cluster of UVA.

\section{Conclusions.}
First principle DFT calculations were employed to scan among semimetallic materials potential candidates  with high Seebeck coefficient. Our calculated Seebeck coefficient values are found to be in agreement with experimental results when the latter were available. A general increase in the intrinsic Seebeck coefficient as a function of materials' bandgap and the ratio of hole mass to electron mass is observed. It is shown that materials with no bandgap but with large mass ratio, can have large Seebeck coefficient values comparable to those of heavily doped semiconductors.
We observed that the Seebeck coefficient values of semimetals were in many cases insensitive to carrier concentration in a wide range around the intrinsic density (see supplementary materials for details). Therefore the coupling between the Seebeck coefficient and the electrical conductivity is weaker in semimetals compared to semiconductors, allowing for simpler optimization of thermoelectric properties.
Many of the studied semimetals including Na$_2$AgSb, Rb$_2$AgSb,TIP, TaP, and HgSe showed Seebeck coefficient values close to or larger than 100 $\mu V/K$.

 Due to relatively high intrinsic carrier concentration and, simultaneously, high mobility, in the absence of doping, these semimetals may show high thermoelectric power factor values. The ones with heavy atoms are good candidates for high zT materials. 

\bibliography{mybiblio}

\end{document}